 \documentclass[aps,prl,twocolumn,groupedaddress,amsmath,amssymb]{revtex4}
\usepackage{graphicx}
\usepackage{amsmath}
\usepackage{dcolumn}% Align table columns on decimal point
\usepackage{bm}% bold math

\begin{document}

%\twocolumn[ %% activate for two-column option

\title{Maximum-likelihood estimation prevents unphysical Mueller matrices}
\author{A. Aiello}
\author{G. Puentes}
\author{D. Voigt}
\author{J.P. Woerdman}
\affiliation{Huygens Laboratory, Leiden University\\
P.O.\ Box 9504, 2300 RA Leiden, The Netherlands}

% Do not use \email or \homepage here. E-mail and URL can be given just before references.

\begin{abstract}
 We show that the method of maximum-likelihood
estimation, recently introduced in the context of quantum process
tomography, can be applied to the determination of  Mueller
matrices characterizing the polarization properties of  classical
   optical systems. Contrary to linear reconstruction
algorithms, the proposed method yields physically acceptable
Mueller matrices even in presence of uncontrolled experimental
errors. We illustrate the method on the case of an unphysical
measured Mueller matrix taken from the literature.

\end{abstract}

\maketitle

% \ocis{000.3860, 260.5430, 290.0290.}

% ] %% activate for two-column option

In the mathematical description of both polarized light and
two-level quantum systems (or \emph{qubits}, in the language of
quantum information), there are
 %one often encounters
  many analogies and common tools. For example, the
Poincar\'{e} sphere \cite{BornWolf} for classical polarization and
the Bloch sphere for two-level quantum systems \cite{Feynman} are,
in fact, the same mathematical object. Although the classical
concepts and tools were introduced well  before the quantum ones,
the latter were developed independently of the former.
Thus, many well established results in classical polarization
optics have been ``rediscovered'' in the context of quantum optics
and quantum information \cite{NielsenBook}. Interestingly, the
inverse process of borrowing results from quantum to classical
optics has started only recently \cite{Legre03a,Brunner03a}.

In this Letter we  give a contribution to this inverse process by
pointing out a connection between \emph{quantum process
tomography} (QPT) \cite{Alte03}  and \emph{classical polarization
tomography} (CPT).
% for linear optical media.
 Specifically, we show that the recently introduced
maximum-likelihood (ML) method for the estimation of quantum
processes \cite{Fiura01a,Sacchi01a,Kosut04a}, can be successfully
applied to the determination of classical Mueller matrices. In the
conventional approach to CPT,  Mueller matrices are estimated from
the measurement data by means of linear algorithms \cite{LeRoy}.
However, such reconstructed Mueller matrices often fail to be
physically acceptable \cite{Anderson94bis}. We show  that this
problem is avoided by using the maximum-likelihood method which
allows to include in a natural manner the
``physical-acceptability'' constraint. Thus, thanks to a ``quantum
tool'', we solve an important issue that has been long debated in
the classical literature \cite{Anderson94,Gopala98a,Gopala98b}.
This is in particular important in view of the present interest in
CPT, e.g., for medical and astronomical imaging.
%%%%%%%%%%%%%%%%%%%%%%%%%%%%%%%%%%%%%%%%%%%%%%%%%%%%%%%%%%%%%

%%%%%%%%%%%%%%%%%%%%%%%%%%%%%%%%%%%%%%%%%%%%%%%%%%%%%%%%%%%%%
%
To begin with, we give first a qualitative description of the
connection between QPT and CPT. At the heart of this connection
lies the well known mathematical equivalence (isomorphism) between
the density matrix $\rho$ describing a two-level quantum system
and the coherency matrix \cite{BornWolf} $J$ describing the
classical polarization state of a light beam
\cite{Falkoff51,Fano54}:
\begin{equation}\label{eq:10}
\rho \sim J/ \mathrm{Tr}\, J .
\end{equation}
$J$ is an Hermitian, positive semidefinite $2 \times 2$ matrix, as
 is $\rho$. A quantum process that transform an input state
$\rho^\mathrm{in}$ in an output state $\rho^\mathrm{out}$ can be
described by a linear superoperator $\mathcal{G}: \,
\rho^\mathrm{out} = \mathcal{G} \rho^\mathrm{in}$. Analogously, a
classical linear optical process (as, e.g., an elastic scattering
process), can be described by a $4 \times 4$  matrix $\mathcal{M}$
such that $J^\mathrm{out} = \mathcal{M} J^\mathrm{in}$ or, in
explicit representation,
\begin{equation}\label{eq:20}
J^\mathrm{out}_{ij} = \sum_{k,l}
\mathcal{M}_{ij,kl}J^\mathrm{in}_{kl}, \qquad i,j,k,l \in \{0,1 \}
.
\end{equation}
In the same way as the reconstruction of $\mathcal{G}$ is the goal
of QPT, the estimation of the elements $\mathcal{M}_{ij,kl}$ from
the measurement data is the goal of CPT. However, in the common
practice, instead of the complex matrix elements
$\mathcal{M}_{ij,kl}$ one wants to determine the $16$ real
elements $M_{\mu \nu}, \, (\mu,\nu=0,\ldots,3)$ of the associated
Mueller matrix $M$.
In the ML approach, the estimated elements of $M$ are found to be
the most likely to yield the measured data. In what follows we
show how to find them.

In a classical polarization tomography experiment, the measurement
data are collected following the scheme shown in Fig. 1.
\begin{figure}[!hb]
\centerline{\includegraphics[width=7cm]{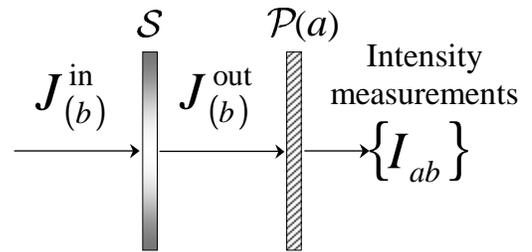}}
\caption{Classical polarization tomography (CPT) scheme.}
\end{figure}
An input light beam is prepared in a pure polarization state
represented by the coherency matrix $J_{(b)}^\mathrm{in}$, and
sent through the optical system $\mathcal{S}$
 where it is transformed in the output beam
represented by
$J_{(b)}^\mathrm{out}=\mathcal{M}J_{(b)}^\mathrm{in}$.
The estimation strategy is to retrieve information on the system
from  a series of polarization tests on the  output states
$J_{(b)}^\mathrm{out}$ obtained from distinct input states
$J_{(b)}^\mathrm{in}, \, (b = 1, \ldots, B)$. A polarization
filter $\mathcal{P}(a)$ that allows the passage of light with
specific polarization labelled by the index $a$, $(a = 1, \ldots,
A)$, provides for the polarization tests.
 Finally, the intensity $I_{ab}$ of the beam after the
filter is recorded. If we prepare $B$ different input states and
we perform $A$ polarization tests per each output state, then such
a CPT experiment will have $A\times B$ outcomes corresponding to
all measured intensities $\{ I_{ab}\},\,( a=1,\ldots,A, \,
b=1,\ldots, B)$.

However, in a ML approach, which is a probabilistic method, one
 deals with relative rather than absolute intensities.
Therefore, from the data set $\{ I_{ab}\}$  we must extract the
relative intensities (or measurement frequencies) $f_{ab} =
I_{ab}/I_{b}$, where $I_{b}$ is the intensity of the input light
beam. Since by definition $0 \leq f_{ab}\leq 1$ and  $ f_{ab} +
f_{a'b}  \leq 1$, where $a,a'$ label two mutually orthogonal
polarization tests, then $f_{ab}$ provides an \emph{experimental}
estimation of the \emph{theoretical} probability $p_{ab}$ for
obtaining a nonzero output intensity with polarization $a$ when
the input beam is prepared in the polarization state $b$.
The theoretical probabilities $\{p_{ab} \}$ can be written in
terms of the input and output states  as \cite{MandelBook}
\begin{equation}\label{eq:30}
p_{ab}(\mathcal{M}) = \frac{\mathrm{Tr}[\Pi_{(a)}
J_{(b)}^\mathrm{out}]}{\mathrm{Tr} \, J_{(b)}^\mathrm{in}},
\end{equation}
where we denoted with $\Pi_{(a)} = \hat{{e}}_{(a)} \otimes
\hat{{e}}_{(a)}^\dagger$ the $2 \times 2$ projection matrix
representing the action of the polarization filter
$\mathcal{P}(a)$ oriented along the (possibly complex) unit vector
$\hat{{e}}_{(a)}$. Since in a CPT experiment  the input beam is
always prepared in a pure polarization state, it can be
represented by a projection matrix $\Pi_{(b)}$ as
$J_{(b)}^\mathrm{in} = I_b \Pi_{(b)}$, where $I_b = \mathrm{Tr} \,
J_{(b)}^\mathrm{in}$ is the intensity of the beam. Then we can
rewrite
\begin{equation}\label{eq:35}
p_{ab}(\mathcal{M})  =  \mathrm{Tr}[\Pi_{(a)} \mathcal{M}
\Pi_{(b)}]\\
\end{equation}
where from now on we assume, without loss of generality, $I_b =1$.
At this point, having measured the frequencies $\{f_{ab} \}$ and
having calculated the probabilities  $\{p_{ab} \}$, the sought
matrix $\mathcal{M}$ can be obtained by maximizing a likelihood
function $\mathcal{L}[\mathcal{M}]$ defined as
\begin{equation}\label{eq:40}
\begin{array}{rcl}
\mathcal{L}[\mathcal{M}] & = & \displaystyle{\sum_{a,b} f_{ab}
\ln[
p_{ab}(\mathcal{M})]}\\
 & = & \displaystyle{\sum_{a,b} I_{ab} \ln\mathrm{Tr}[\Pi_{(a)} \mathcal{M}
\Pi_{(b)}]},
\end{array}
\end{equation}
 where $p_{ab}(\mathcal{M}) \geq 0$ for any physically acceptable
 process.

 Equation (\ref{eq:40}) is the first main result of this
 paper. It contains both experimental ($I_{ab}$) and theoretical
 [$p_{ab}(\mathcal{M})$] quantities.
 Now, we demonstrate that it is possible to impose the condition $p_{ab}(\mathcal{M}) \geq 0$
\emph{before} the maximization operation, in such a way that the
estimated Mueller matrix is automatically physically acceptable.
After a lengthy but straightforward calculation, it is possible to
show that the matrix $\mathcal{M}$ can be written in terms of an
Hermitian matrix $H$ as \cite{Aiello04d}
\begin{equation}\label{eq:50}
\mathcal{M}=\sum_{\mu,\nu}^{0,3} H_{\mu \nu} \left[
\epsilon_{(\mu)} \otimes \epsilon_{(\nu)} \right],
\end{equation}
where $\{\epsilon_{(\mu)}\}$ are the elements of the standard
basis in $\mathbf{C}^{2 \times 2}$. By substituting Eq.
(\ref{eq:50}) into Eq. (\ref{eq:30}) we obtain
\begin{equation}\label{eq:60}
p_{ab}(\mathcal{M})=\sum_{\mu,\nu}^{0,3} H_{\mu \nu} \mathrm{Tr} [
\Pi_{(a)} \epsilon_{(\mu)} \Pi_{(b)}\epsilon_{(\nu)}^T ],
\end{equation}
where the superscript $T$ indicates the transposed matrix. The
probabilities $p_{ab}(\mathcal{M})$ as written in Eq.
(\ref{eq:60}) can still be negative, because only  $H$ matrices
associated with physically acceptable Mueller matrices can
guarantee the condition $p_{ab} \geq 0 $. However, we know from
the Mueller matrix theory
 that the $H$ matrix associated with a physically acceptable
Mueller matrix must be positive semidefinite \cite{Anderson94}. It
is well known that any positive semidefinite matrix can be written
in terms of its  Cholesky decomposition as
\begin{equation}\label{eq:70}
H = 2\frac{C C^\dagger}{\mathrm{Tr}(C C^\dagger) },
\end{equation}
where $C$ is a lower triangular matrix
\begin{equation}\label{eq:80}
C = \begin{pmatrix}
  h_{1} & 0 & 0 & 0 \\
  h_{5} + \mathrm{i} h_{6} & h_{2} & 0 & 0 \\
  h_{11} + \mathrm{i} h_{12} & h_{7} + \mathrm{i} h_{8} & h_{3} & 0 \\
  h_{15} + \mathrm{i} h_{16} & h_{13} + \mathrm{i} h_{14} & h_{9} + \mathrm{i} h_{10} & h_{4}
\end{pmatrix},
\end{equation}
composed by $16$ real parameters $h_k, \, (k =1, \ldots, 16)$, and
we fixed the normalization of $H$ by setting $M_{00}=1$. Then,
after substituting Eq. (\ref{eq:70}) into Eq. (\ref{eq:60}), the
maximum of $\mathcal{L}$ can be found by using a standard
maximization algorithm \cite{RecipeFor}. The search for the
maximum is performed in the real $16$-dimensional space of
parameters $\{ h_k \}$. Once the optimal set of values
$\{h_1^\mathrm{opt}, \ldots, h_{16}^\mathrm{opt} \}$ that maximize
$\mathcal{L}$ has been found, this  can be used in Eq.
(\ref{eq:70}) to obtain the corresponding $H^\mathrm{opt}$.
Finally, the elements of the sought physically acceptable
 Mueller matrix can be computed as
\begin{equation}\label{eq:90}
M_{\mu \nu} = \mathrm{Tr} \left\{H^\mathrm{opt} [\sigma_{(\mu)}
\otimes \sigma_{(\nu)}^*]\right\},
\end{equation}
where $\{\sigma_{(\mu)} \}$ are the normalized Pauli matrices
\cite{Aiello04d}. This is our second main result. A Mueller matrix
$M$ determined in this way represents the answer to the question:
``\emph{which physically acceptable Mueller matrix is most likely
to yield the measured data? }''

The rest of the paper is devoted to the illustration of the theory
outlined above, by applying it to a realistic case. We have chosen
from the current literature \cite{Howell}  the following Mueller
matrix which was already shown \cite{Anderson94} to be physically
unacceptable:
\begin{equation}\label{eq:100}
M' =
\begin{pmatrix}
 {0.7599} & {0.0295} & {0.1185} &
   -\text{0.0623} \\
 {0.0384} & {0.5394} & {0.0282} &
   -\text{0.1714} \\
 {0.124} & -\text{0.012} & {0.6608} &
   {0.2168} \\
 -\text{0.0573} & -\text{0.1811} &
   -\text{0.1863} & {0.4687}
\end{pmatrix}.
\end{equation}
From $M'$ we calculated the (normalized) associated Hermitian
matrix $H'$ which is \emph{not} positive semidefinite since it has
one negative eigenvalue:
\begin{equation}\label{eq:105}
\mathrm{diag} \, H' = \{1.6671, \, 0.2950, \, 0.2330, \, -0.1951
\}.
\end{equation}
By using Eq. (\ref{eq:60}), we generated
 a set of $36$ ``fake-measured'' data $f_{ab}'$ as
\begin{equation}\label{eq:110}
f_{ab}'=\sum_{\mu,\nu}^{0,3} H_{\mu \nu}' \mathrm{Tr} [ \Pi_{(a)}
\epsilon_{(\mu)} \Pi_{(b)} \epsilon_{(\nu)}^T ],
\end{equation}
where we selected both the input beam (represented by $\Pi_{(b)}$)
and the polarization filter (represented by  $\Pi_{(a)}$) from the
set of $6$  pure polarization states  labelled as horizontal (H)
and  vertical (V), oblique at $45^\circ$ and oblique at
$135^\circ$, right (RHC) and left (LHC). The so obtained $36$
numbers represent our ``experimental'' data set. Obviously, from
these numbers  one could generate back $M'$ with the conventional
linear algorithm. It may worth to note that the three pairs of
polarization states we have chosen, are the ones usually employed
in CPT \cite{BornWolf}, and correspond to three mutually unbiased
basis \cite{PeresBook} often utilized in QPT. We used the
MATHEMATICA  5.1 function FindMaximum to maximize $\mathcal{L}$.
Tho run this function it is necessary to furnish a set of initial
values for the parameters $\{ h_1,\ldots,h_{16} \}$ to be
estimated. We found convenient to proceed in the following way: we
made  first a polar decomposition of $H'$ to obtain the positive
semidefinite matrix $H'' = 2\sqrt{H'
{H'}^\dagger}/\mathrm{Tr}\sqrt{H' {H'}^\dagger}$, then we obtain
the initial values from the Cholesky decomposition of $H''$.
Finally, after maximization, we obtained the maximum-likelihood
estimation of $M^\mathrm{ML}$ as
\begin{equation}\label{eq:120}
M^\mathrm{ML}=\begin{pmatrix}
 \text{0.7599} & \text{0.0257} &
   \text{0.1206} &
   -\text{0.0576} \\
 \text{0.0372} &
   \text{0.5285} &
   \text{0.0001} &
   -\text{0.0496} \\
 \text{0.1208} &
   -\text{0.0001} &
   \text{0.6184} &
   \text{0.1920} \\
 -\text{0.0554} &
   -\text{0.0572} &
   -\text{0.1794} &
   \text{0.4822}
\end{pmatrix}.
\end{equation}
As expected, this matrix is indeed a physically acceptable Mueller
matrix, as the eigenvalues of its associated $H^\mathrm{ML}$
matrix are all non-negative:
\begin{equation}\label{eq:125}
\mathrm{diag} \,H^\mathrm{ML} = \{1.6344,\,  0.2341, \, 0.1315,
\,0\}.
\end{equation}
A visual inspection show that $M$ and $M^\mathrm{ML}$ differs only
by a little amount. This was expected since we choose an initial
Mueller matrix $M'$ that is not very unphysical (only one negative
eigenvalue ). A quantitative estimation of the  difference between
$M'$ and $M^\mathrm{ML}$ can be given by calculating their
relative Frobenius distance \cite{LeRoy}
\begin{equation}\label{eq:130}
\frac{||M -M^\mathrm{ML}||}{||M +M^\mathrm{ML}||}= 0.072 ,
\end{equation}
which indicates that the average relative difference between
corresponding matrix elements of $M$ and $M^\mathrm{ML}$ is about
$7 \%$. This confirms the quality of our approach even with sparse
data set (only 36 values).

In conclusion, we have shown that it is possible to apply the
maximum-likelihood method, initially developed for quantum process
tomography, to the classical problem of  Mueller matrix
reconstruction. Moreover, we have shown that  this method  has the
benefit to produce \emph{always} physically acceptable Mueller
matrices as the most likely matrices which yield the measured
data.

We acknowledge support from the EU under the IST-ATESIT contract.
This project is also supported by FOM.


\begin{thebibliography}{10}
\bibitem{Anderson94bis}
In the current literature there is some confusion in the use of
terms such as  ``nominal'' Mueller matrix, ``physical'' Mueller
matrix, etc.. Here we adopt   the terminology of Ref.
\cite{Anderson94}.
\end{thebibliography}
\end{document}